\documentclass{aiaa-pretty}

\usepackage{lettrine}
\usepackage{mathtools}
\usepackage{amsmath}
\usepackage{float}

\author[Tyson and Roy]{ %
William C. Tyson\thanks{Graduate Research Assistant,
Department of Aerospace and Ocean Engineering, 215 Randolph Hall,
AIAA Student Member} and
Christopher J. Roy\thanks{Professor,
Department of Aerospace and Ocean Engineering, 215 Randolph Hall,
AIAA Associate Fellow} \\
\textit{Virginia Tech, Blacksburg, VA 24061}}

\title{Efficient Functional-Based Adaptation for CFD Applications}

\abstract{
Adjoint methods have gained popularity in recent years for driving adaptation
procedures which aim to reduce error in solution functionals.
While adjoint methods have been proven effective for functional-based
adaptation, the practical implementation of an adjoint method can be quite
burdensome since code developers constantly need to ensure and maintain a dual
consistent discretization as updates are made.
Also, since most engineering problems consider multiple functionals, an adjoint
solution must be obtained for each functional of interest which can increase
the overall computational cost significantly.
In this paper, an alternative to adjoints is presented which uses a sparse
approximate inverse of the Jacobian of the residual to obtain approximate
adjoint sensitivities for functional-based adaptation indicators.
Since the approximate inverse need only be computed once, it can be recycled
for any number of functionals making the new approach more efficient than a
conventional adjoint method.
This new method for functional-based adaptation will be tested using the
quasi-1D nozzle problem, and results are presented for functionals of
integrated pressure and entropy.

}

\begin{document}
\maketitle

\section{Introduction}
\lettrine{W}{ith} the onset of modern computing, computational fluid
dynamics (CFD) has become an integral part of both engineering design and
analysis.
In its short history, CFD has proven to be a very powerful tool in the
development and optimization of engineering devices.
Although it can offer a tremendous amount of insight into the flow physics
governing a particular problem, there will always be some amount of error
inherent in CFD simulations due to the required discretization of the
governing equations.
The presence of this inherent error can be easily forgotten if CFD simulations
are viewed as ``truth'' rather than as an approximation of the true physics.
Also, as engineers attempt to optimize their designs to balance performance,
efficiency, and cost, the success or failure of their device can hinge on
the accuracy of a simulation.
Therefore, it is becoming increasingly more important to be able to accurately
assess and reduce the numerical error that is present in CFD simulations.

Numerical error is introduced into a simulation through various aspects of the
solution procedure such as the chosen discretization scheme and geometry
definition.
Errors are typically categorized as one of the following: round-off error,
iterative error, or discretization error.
Round-off error and iterative error are generally not the leading contributors
to the overall error in a solution.
Due to the increased precision modern computers, round-off error is usually
small enough to be neglected.
Iterative error will also be small enough to be nglected as long as numerical
solutions are fully converged.
Discretization error, which is the difference between the exact solution to
the discrete governing equations and the exact solution to the continuous
governing equations, is often the primary source of numerical error.
Accurately assessing the amount of discretization error in a solution can be
a very difficult problem; while the discrete numerical solution can serve as
the exact solution to the discrete equations to within round-off and iterative
error, the exact solution to the continuous governing equations is rarely known.
As a result, much research has been conducted to develop methods which can
accurately estimate discretization error~\cite{Roy2010}, the most common of
which are mesh refinement and residual-based methods.

Mesh refinement methods for discretization error estimation, such as
Richardson extrapolation~\cite{Richardson1911}, require numerical solutions on
multiple grid levels in order to obtain a discretization error estimate.
Since only the numerical solution is needed, implementation of a mesh
refinement method is straightforward and requires minimal code modifications to
produce an error estimate.
While implementation may be relatively cheap, a major requirement for these
methods is that all grid levels used in obtaining an error estimate must be in
the asymptotic range~\cite{Roy2010}.
For a solution to be in the asymptotic range, the mesh resolution must be fine
enough such that all higher order terms are small and the discretization error
converges at the formal order of accuracy of the discretization.
Obtaining solutions in the asymptotic range can be quite difficult especially
for nonsmooth problems.
Increasing the resolution of the grid uniformly everywhere in order to achieve
the asymptotic range can become very computationally expensive.
When simulations can sometimes take days or even weeks to run for large, 3D
engineering problems, mesh refinement methods for discretization error
estimation can quickly become impractical.

Residual-based methods were developed as a cheaper alternative to mesh
refinement methods.
With a residual-based method, the numerical solution from just one grid level
can be used in conjuction with the residual to formulate a discretization
error estimate.
The reliability of these types of discretization error estimators is heavily
dependent upon accurately representing the truncation error, which is the
difference between the continuous governing equations and the discrete
governing equations operating on the same solution.
For complex partial differential equations, deriving an analytic representation
of the truncation error, if possible, is a very time-consuming and error-prone
process.
A numerical truncation error representation is usually constructed in practice.
Phillips et al.~\cite{Phillips2013} have had much success numerically
estimating truncation error for manufactured solutions to the Euler equations.
Once the truncation error is determined, an estimate of discretization error
can be established everywhere in the domain.

Equally important to assessing the amount of error in a solution is being able
to reduce it.
Discretization error can be reduced with uniform grid refinement, but, due to
the exponential increase in computational cost, is generally not a viable
option for large engineering problems.
For instance, in 3D, the computational cost increases on the order of
r\textsuperscript{3} where r is the grid refinement factor.
Adaptation, on the other hand, provides an efficient mechanism for increasing
solution accuracy without exponentially increasing computation time.
Prior to conducting adaptation, two choices must be made which will dictate the
performance of the adaptation process: 1) the adaptation indicator and 2) the
type of adaptation.
The adaptation indicator acts as a flag to target particular cells where
adaptation needs to be performed and is used to drive the chosen type of
adaptation.

The selection of an appropriate adaptation indicator is the primary factor
governing how the adaptation behaves and ultimately the extent to which
discretization error will be reduced.
Feature-based adaptation flags cells for adaptation based upon the gradient or
curvature of one or more flowfield variables making it attractive for
resolving shocks and other discontinuous flow features.
While feature-based adaptation can provide sharp resolution of discontinuities,
this type of adaptation is not guaranteed to reduce numerical errors and can
even introduce more error into the solution~\cite{Warren1991,AinsworthOden2000}
because it does not account for how discretization error is generated,
transported, and dissipated.
Rather, feature-based adaptation assumes that error is localized at the
feature which may not necessarily be the case.
This heuristic approach to adaptation can lead to under-resolution in areas of
the domain which contribute most to the discretization error and
over-resolution in areas which contribute very little.
Truncation error based adaptation seeks to remedy this deficiency on the part
of feature-based adaptation.
Truncation error, which will be presented in a later section, acts as the
local source of discretization error in numerical
solutions~\cite{Roy2009,Oberkampf2010}.
By targeting regions of the domain with high truncation error, the local
production of discretization error can be reduced.
Choudhary and Roy~\cite{Choudhary2012} have had success applying this type of
adaptation to 1D and 2D Burgers equation.
Tyson et al.\cite{Tyson2015} extended their work in 2D to investigate the
application of different adaptation schemes for truncation error based
adaptation and were able to achieve substantial reductions in discretization
error equivalent to three levels of uniform refinement.
While truncation error based adaptation does target the local source of
discretization error, like feature-based adaptation, it does not account for
how discretization error is transported and diffused throughout the domain.

Adjoint methods have emerged as the state of the art for adaptation which aims
to increase the accuracy of a discrete solution functional.
These methods have gained popularity due to their rigorous mathematical
foundation and because the solution to the adjoint, or dual, problem can also be
used to formulate an error estimate in the functional of interest.
Adjoint-based adaptation targets areas of the domain which contribute to the
overall error in the functional.
Venditti and Darmofal~\cite{Venditti2000} applied adjoint methods to
quasi-1D flows using an embedded grid technique where a reconstructed coarse
grid solution was restricted to an embedded fine grid.
In this way, they were able to obtain an error estimate in the solution
functional on a fine grid using a coarse grid solution.
The solution to the adjoint problem was then used to adapt the coarse grid to
obtain a better error estimate on the fine grid.
Venditti and Darmofal have also had much success extending their work to 2D
inviscid and viscous flows~\cite{Venditti2002,Venditti2003}.
While adjoint methods have shown great promise for both error estimation and
adaptation, the solution to the adjoint problem only supplies an error
estimate and adaptivity information for one functional.
Typically, engineers require information about multiple functionals
such as three forces and three moments for 3D aerodynamics problems.
Derlaga~\cite{Derlaga2015b} was able to circumvent this issue for functional
error estimation by showing that local residual-based error estimates can
provide the same functional error estimate as an adjoint solution when the
same linearization and truncation error estimate are used.
This allows error estimates for multiple functionals to be obtained at little
to no cost compared to an adjoint method.
But, for functional-based adaptation, Park et al.~\cite{Park2011} and
Fidkowski and Roe~\cite{Fidkowski2009} have found that adaptation based solely
on the adjoint solution for one functional can actually increase the error in
another functional.
Therefore, for functional-based adaptation, it is still necessary to solve
multiple adjoint problems in order to improve all functionals of interest.
Solving multiple adjoint problems can become quite costly as the number of
functionals increases both in terms of storage and computational time.
This could ultimately hinder the widespread use of adjoint methods in
commercial applications.

Since most engineering design choices are based upon a given functional of the
solution, it is imperative that methods for obtaining functional error
estimates and adaptivity information be not only accurate but also efficient.
In this paper, we seek an efficient alternative to adjoint methods for
functional-based adaptation.
First, we will present the relationship between truncation error and
discretization error.
Adjoint methods for error estimation and adaptation will be reviewed followed
by our proposed alternative to adjoints for functional-based adaptation.
The new method for functional-based adaptation will be tested using the
quasi-1D Euler equations, and adaptation indicators will be compared with
those generated by a conventional adjoint solution.
Adaptation will be performed and functional discretization error improvements
for both methods will be presented.

\section{Truncation Error \& Discretization Error}

\subsection{Generalized Truncation Error Expression}
Before the relationship between truncation error and discretization error can
be presented, the relationship between the continuous form and the discrete
form of the governing equations must be understood.
A given set of continuous governing equations, $L(\cdot)$, may be related to
a consistent discretized form of the equations, $L_h(\cdot)$, by the
Generalized Truncation Error Expression (GTEE)~\cite{Oberkampf2010}:

\begin{equation}\label{GTEE1}
L_{h}(I^{h}u) = I^{h}L(u) + \tau_{h}(u)
\end{equation}

\noindent where $u$ is any continuous function and $\tau_{h}(\cdot)$
represents the truncation error.
From Eq.~\ref{GTEE1}, it can be seen that the truncation error, as the
difference between the continuous and discrete forms of the governing
equations, represents higher order terms which are truncated in the
discretization process.
For model problems, the truncation error can be shown to be a function of
continuous solution derivatives and cell size~\cite{Choudhary2011}.
But, while operating on a continuous solution space, evaluating the truncation
error results in a discrete space~\cite{Phillips2013,Phillips2011}.
The operator, $I_{a}^{b}$, is a restriction or prolongation operator which
allows for the transition between a continuous space and a discrete space;
the subscript, $a$, denotes the starting space and the superscript, $b$,
denotes the resultant space.
For the GTEE, a subscript or superscript $h$ denotes a discrete space on a
mesh with a characteristic size, $h$, and an empty subscript or superscript
represents a continuous space.
Also, there is no restriction on the form of the continuous and discrete
operators, $L(\cdot)$ and $L_{h}(\cdot)$; the governing equations can be
strong form (ODEs and PDEs) or weak form (integral equations).

\subsection{Error Transport Equations}

Having related the contiuous governing equations, $L(\cdot)$, to the discrete
governing equations, $L_{h}(\cdot)$, through the truncation
error, $\tau_{h}(\cdot)$, we are now in a position to relate truncation
error to the discretization error.
First, consider the following definitions.
For the exact solution to the continuous governing equations, $\tilde{u}$,
and the exact solution to the discrete governing equations, $u_{h}$, the
following holds:
\begin{subequations}
\begin{align}
L(\tilde{u}) &= 0 \label{soln_def1} \\
L_{h}(u_{h}) &= 0 \label{soln_def2}
\end{align}
\end{subequations}

\noindent Plugging the exact solution to the discrete equations, $u_{h}$,
into the GTEE and noting that the left hand side is
identically zero by Eq.~\ref{soln_def2}, the GTEE may be simplified to the
following:

\begin{equation}\label{GTEE2}
0 = I^{h}L(I_{h}u_{h}) + \tau_{h}(I_{h}u_{h})
\end{equation}

\noindent where $I_{h}$ prolongs the exact discrete solution to a continuous
space.
Next, the continuous governing equations, Eq.~\ref{soln_def1}, can be
subtracted from both sides of Eq.~\ref{GTEE2} to form:

\begin{equation}\label{GTEE3}
0 = I^{h}L(I_{h}u_{h}) - I^{h}L(\tilde{u}) + \tau_{h}(I_{h}u_{h})
\end{equation}

\noindent Finally, if we define the discretization error as the difference
between the exact solution to the discrete equations and the exact solution
to the continuous equations, $I_{h}\varepsilon_{h} = I_{h}u_{h}-\tilde{u}$, and
if the governing equations are linear or
linearized~\cite{Phillips2011,Oberkampf2010}, the continuous operators in
Eq.~\ref{GTEE3} can be combined and the definition of discretization error may
be inserted to form the continuous form of the error transport equations (ETE):

\begin{equation}\label{continuous_ETE}
I^{h}L(I_{h}\varepsilon_{h}) = -\tau_{h}(I_{h}u_{h})
\end{equation}

\noindent An equivalent discrete form of the ETE can also be derived and are
given by:

\begin{equation}\label{discrete_ETE}
L_{h}(\varepsilon_{h}) = -\tau_{h}(\tilde{u})
\end{equation}

\noindent The ETE provide a great deal of information regarding how
discretization error behaves and how it is related to the truncation error.
Eq.~\ref{continuous_ETE} and Eq.~\ref{discrete_ETE} demonstrate that
truncation error acts as the local source for discretization error in numerical
solutions and that discretization error is convected and diffused in the same
manner as the solution.

\subsection{Truncation Error Estimation}

In practice, the discrete form of the ETE, Eq.~\ref{discrete_ETE}, are used to
solve for the local discretization error throughout the domain.
But in order to do this, the exact solution to the continuous
equations, $\tilde{u}$, must be known to evaluate the truncation error.
Since the exact continuous solution is in practice never known, the truncation
error is typically estimated with a reconstruction of the discrete solution,
that is:

\begin{equation}\label{estimated_TE1}
\tau_{h}(\tilde{u}) \approx \tau_{h}(I_{h}^{q}u_{h})
\end{equation}

\noindent where $I_{h}^{q}u_{h}$ is a $q^{th}$-order reconstruction of the
exact discrete solution, $u_{h}$.
The discrete numerical solution is often a very good approximation of the
exact discrete solution, $u_{h}$, assuming round-off and iterative errors are
small.
To estimate the truncation error, the exact solution to the discrete equations
is inserted into Eq.~\ref{GTEE1} such that~\cite{Phillips2013}:

\begin{equation}\label{estimated_TE2}
\tau_{h}(I_{h}^{q}u_{h}) = -I^{h}L(I_{h}^{q}u_{h}) + O(h^{\bar{q}})
\end{equation}

\noindent where $\bar{q}$ is an order similar to $q$ but altered by the
continuous govering equations, $L(\cdot)$.
Therefore, operating the continuous residual, $L(\cdot)$, on the reconstructed
solution provides the necessary truncation error estimate for solving the ETE.

\section{Functional-Based Adaptation}
\subsection{Adjoint Methods}

While adjoint methods were originally used in design optimization to obtain
sensitivities of a functional to a set of design parameters~\cite{Jameson1988},
adjoint methods have gained much popularity within the CFD community for
obtaining functional error estimates and adaptation indicators.
Solutions from the adjoint, or dual, problem can be used to drive an
adaptation procedure which only targets regions of the domain which contribute
to the discretization error in the functional of interest.
In order to illustrate why adjoint methods can be used in this way, first
consider a Taylor expansion of the discrete equations, $L_{h}(u_{h})$, about a
general discrete function, $u$:

\begin{equation}\label{residual_expansion}
L_{h}(u_{h}) = L_{h}(u) + \frac{\partial L_{h}}{\partial U}\bigg|_{U=u}(u_{h}-u) +
\frac{\partial^{2}L_{h}}{\partial U^{2}}\bigg|_{U=u}\frac{(u_{h}-u)^{2}}{2} +
\ldots
\end{equation}

\noindent Likewise, consider a similar expansion of a discrete solution
functional, $J_{h}$, about the same general discrete function, $u$:

\begin{equation}\label{functional_expansion1}
J_{h}(u_{h}) = J_{h}(u) + \frac{\partial J_{h}}{\partial U}\bigg|_{U=u}(u_{h}-u) +
\frac{\partial^{2}J_{h}}{\partial U^{2}}\bigg|_{U=u}\frac{(u_{h}-u)^{2}}{2} +
\ldots
\end{equation}

\noindent Combining Eq.~\ref{residual_expansion} and
Eq.~\ref{functional_expansion1} in a Lagrangian, the functional,
$J_{h}(\cdot)$, evaluated at the exact discrete solution, $u_{h}$, may be
rewritten as:

\begin{equation}\label{adjoint_lagrangian1}
\begin{aligned}
J_{h}(u_{h}) &= \left[ J_{h}(u) +
\frac{\partial J_{h}}{\partial U}\bigg|_{U=u}(u_{h}-u) +
\frac{\partial^{2}J_{h}}{\partial U^{2}}\bigg|_{U=u}\frac{(u_{h}-u)^{2}}{2} +
\ldots \right] \\
+& \lambda ^{T}\left[ L_{h}(u) +
\frac{\partial L_{h}}{\partial U}\bigg|_{U=u}(u_{h}-u) +
\frac{\partial^{2}L_{h}}{\partial U^{2}}\bigg|_{U=u}\frac{(u_{h}-u)^{2}}{2} +
\ldots \right]
\end{aligned}
\end{equation}

\noindent where $\lambda^{T}$ is a 1xN vector of Lagrange multipliers where N
is the total number of unknowns.
This operation is made possible by the fact that the terms in brackets
following the Lagrange multipliers, $\lambda$, are exactly equal to zero by
Eq.~\ref{residual_expansion} and Eq.~\ref{soln_def2}.
Now, neglecting higher order terms, inserting a restriction of the exact
solution to the continuous equations, $I^{h}\tilde{u}$, for the general discrete
function, $u$, and using the definition of discretization error,
$\varepsilon_{h} = u_{h} - I^{h}\tilde{u}$, Eq.~\ref{adjoint_lagrangian1} may be
simplified to the following:

\begin{equation}\label{adjoint_lagrangian2}
J_{h}(u_{h}) \approx J_{h}(I^{h}\tilde{u}) +
\frac{\partial J_{h}}{\partial U}\bigg|_{U=I^{h}\tilde{u}}\varepsilon_{h} +
\frac{\partial^{2}J_{h}}{\partial U^{2}}\bigg|_{U=I^{h}\tilde{u}}
\frac{\varepsilon_{h}^{2}}{2} + \lambda^{T} \left[ L_{h}(I^{h}\tilde{u}) +
\frac{\partial L_{h}}{\partial U}\bigg|_{U=I^{h}\tilde{u}}\varepsilon_{h} +
\frac{\partial^{2}L_{h}}{\partial U^{2}}\bigg|_{U=I^{h}\tilde{u}}
\frac{\varepsilon_{h}^{2}}{2}
\right]
\end{equation}

\noindent If the first term on the right hand side is taken to the left hand
side and the error in the functional is approximated as
$\varepsilon_{J} = J_{h}(u_{h}) - J_{h}(I^{h}\tilde{u})$,
Eq.~\ref{adjoint_lagrangian2} becomes analogous to solving a constrained
optimization problem which seeks to minimize the error in the functional
subject to satisfying the discrete governing equations, $L_{h}(u_{h}) = 0$.
Finally, if the exact solution to the continuous governing
equations, $\tilde{u}$, is inserted into the GTEE so that
$L_{h}(I^{h}\tilde{u}) = \tau_{h}(\tilde{u})$, Eq.~\ref{adjoint_lagrangian2} may
be simplified and rearranged to form:

\begin{equation}\label{functional_error1}
\underbrace{\vphantom{
\left[\frac{\partial J_{h}}{\partial U}\bigg|_{U=I^{h}\tilde{u}}\right]}
J_{h}(u_{h}) - J_{h}(I^{h}\tilde{u})}_\text{Functional Error}
\approx \underbrace{\vphantom{
\left[\frac{\partial J_{h}}{\partial U}\bigg|_{U=I^{h}\tilde{u}}\right]}
\lambda^{T} \tau_{h}(\tilde{u})}_\text{Correction} +
\varepsilon_{h} \underbrace{\left[
\frac{\partial J_{h}}{\partial U}\bigg|_{U=I^{h}\tilde{u}} +
\lambda^{T} \frac{\partial L_{h}}{\partial U}\bigg|_{U=I^{h}\tilde{u}}
\right]}_\text{Adjoint Problem} +
\underbrace{ \frac{\varepsilon_{h}^{2}}{2} \left[
\frac{\partial^{2}J_{h}}{\partial U^{2}}\bigg|_{U=I^{h}\tilde{u}} +
\lambda^{T} \frac{\partial^{2}L_{h}}{\partial U^{2}}\bigg|_{U=I^{h}\tilde{u}}
\right]}_\text{Remaining Error}
\end{equation}

\noindent Eq.~\ref{functional_error1} represents the discrete adjoint
formulation of the functional error.
An equivalent continuous formulation can be derived in a similar manner.
Before moving forward, it is important to note that from now on the
Lagrange multipliers, $\lambda$, will be referred to as the adjoint variables
or adjoint sensitivities.
The adjoint variables represent the sensitivity of the functional to local
perturbations of the residual.

The functional error estimate consists of three parts: (1) the computable
adjoint correction, (2) the solution of the adjoint problem, and (3) higher
order remaining error terms.
The adjoint correction is computed by first solving the adjoint problem
such that the terms in brackets are equal to zero:

\begin{equation}\label{adjoint_problem1}
\frac{\partial J_{h}}{\partial U}\bigg|_{U=I^{h}\tilde{u}} +
\lambda^{T} \frac{\partial L_{h}}{\partial U}\bigg|_{U=I^{h}\tilde{u}} = 0
\end{equation}

\noindent Rearranging Eq.~\ref{adjoint_problem1} yields a system of linear
equations for the adjoint variables, $\lambda$:

\begin{equation}\label{adjoint_problem2}
\left[ \frac{\partial L_{h}}{\partial U}\bigg|_{U=I^{h}\tilde{u}} \right]^{T}
\lambda =
- \left[ \frac{\partial J_{h}}{\partial U}\bigg|_{U=I^{h}\tilde{u}} \right]^{T}
\end{equation}

\noindent It is important to note that the term in brackets on the left hand
side of Eq.~\ref{adjoint_problem2} is the Jacobian of the primal
solution which, for a code that includes an implicit solver, will already
be computed and stored.
Already having computed the Jacobian can make implementation of an adjoint
method slightly less challenging.
Now, upon inspection of Eq.~\ref{adjoint_problem2}, it can be seen that
the adjoint variables represent the discrete sensitivities of the functional
to perturbations in the discrete operator making them an ideal candidate for
flagging areas of the domain for adaptation which contribute most to error in
the chosen functional.
Once the adjoint problem has been solved for the adjoint variables, the
adjoint correction term may be computed assuming an accurate truncation error
estimate has been determined such that Eq.~\ref{estimated_TE1} holds.
Finally, having solved the adjoint problem and having computed the adjoint
correction, all that remains are higher order remaining error terms.
These higher order terms can be quite expensive to compute and are most often
left unquantified.
But, for a second order accurate discretization, while the adjoint correction
reduces at a second order rate, the remaining error terms will drop at a
fourth order rate.
This higher order convergence of the remaining error terms indicates that even
for coarse grids the magnitude of the remaining error can be small relative
to the adjoint correction.

\subsection{Approximate Adjoint Methods}

Adjoint methods appear to be ideal for functional-based error estimation and
adaptation because of their rigorous mathematical foundation and because both
error estimates and adaptation indicators can be obtained in one compact
procedure.
But, since most engineering problems require analyzing multiple solution
functionals, the adjoint problem must be solved for each functional of
interest.
As the number of required functionals increases, the cost of solving several
adjoint problems can become impractical.
Research has also shown that adaptation which solely targets one functional
can actually increase the error in another
functional~\cite{Park2011,Fidkowski2009} thus furthering the need to solve
the adjoint problem for all functionals of interest.
In addition to an increase in computational cost, proper implementation of an
adjoint solver can be quite burdensome.
The developer must pay careful consideration to maintaining a dual consistent
discretization between the primal solver and the adjoint solver so that
information is not transported in a manner inconsistent with the physics of the
problem.
Derlaga et al.~\cite{Derlaga2013} offer an example of how a dual inconsistent
implementation of boundary conditions can corrupt the adjoint solution.

While adjoint methods have been proven effective for functional error
estimation and adaptation, these methods have clear disadvantages which could
ultimately prevent their widespread use in commercial applications.
In this section, we present an alternative approach to adjoint methods whose
sole purpose is to increase the efficiency of functional-based adaptation by
alleviating the need to solve an adjoint problem for each functional of
interest.
To accomplish this, we must first derive the implicit form of the discrete ETE,
Eq.~\ref{discrete_ETE}.
By inserting the exact solution to the continuous governing
equations, $\tilde{u}$, into the GTEE and subtracting Eq.~\ref{soln_def2}
from both sides, the GTEE can be rewritten as:

\begin{equation}\label{GTEE4}
L_{h}(u_{h}) - L_{h}(I^{h}\tilde{u}) = -\tau_{h}(\tilde{u})
\end{equation}

\noindent Next, consider an expansion of the discrete governing
equations, $L_{h}(\cdot)$, operating on a restriction of the exact continuous
solution, $I^{h}\tilde{u}$, about the exact discrete solution, $u_{h}$:

\begin{equation}\label{linearized_residual}
L_{h}(I^{h}\tilde{u}) = L_{h}(u_{h}) -
\frac{\partial L_{h}}{\partial U}\bigg|_{U=u_{h}} \varepsilon_{h} +
O(\varepsilon_{h}^{2})
\end{equation}

\noindent Inserting Eq.~\ref{linearized_residual} and Eq.~\ref{estimated_TE1}
into Eq.~\ref{GTEE4} and
neglecting higher order terms, we arrive at the implicit form of the discrete
ETE:

\begin{equation}\label{linearized_ETE}
\frac{\partial L_{h}}{\partial U}\bigg|_{U=u_{h}} \varepsilon_{h} \approx
-\tau_{h}(I_{h}^{q}u_{h})
\end{equation}

\noindent where the $\frac{\partial L_{h}}{\partial U}\bigg|_{U=u_{h}}$ is simply
the Jacobian for the primal solution which again for a code that implements an
implicit primal solver is already available.
Now, consider a similar expansion of the functional about the exact solution
to the discrete equations, $u_{h}$:

\begin{equation}\label{functional_expansion2}
J_{h}(I^{h}\tilde{u}) = J_{h}(u_{h}) +
\frac{\partial J_{h}}{\partial U}\bigg|_{U=u_{h}}
\underbrace{(u_{h}-I^{h}\tilde{u})}_{\varepsilon_{h}} + O(\varepsilon_{h}^{2})
\end{equation}

\noindent Noting that the term in parenthesis is the definition of
discretization error, $\varepsilon_{h}$, and by rearranging and neglecting
higher order terms, the error in the functional can be expressed as:

\begin{equation}\label{functional_error2}
\varepsilon_{J} = J_{h}(u_{h}) - J_{h}(I^{h}\tilde{u}) \approx
-\frac{\partial J_{h}}{\partial U}\bigg|_{U=u_{h}} \varepsilon_{h}
\end{equation}

\noindent Solving Eq.~\ref{linearized_ETE} for the discretization
error, $\varepsilon_{h}$, and inserting into Eq.~\ref{functional_error2},
the error in the functional can be written as the following:

\begin{equation}\label{functional_error3}
\varepsilon_{J} \approx \frac{\partial J_{h}}{\partial U}\bigg|_{U=u_{h}}
\left[ \frac{\partial L_{h}}{\partial U}\bigg|_{U=u_{h}} \right]^{-1}
\tau_{h}(I_{h}u_{h})
\end{equation}

\noindent By comparing Eq.~\ref{functional_error3} with
Eq.\ref{functional_error1}, it can be seen that if the
Jacobian, $\frac{\partial L_{h}}{\partial U}\bigg|_{U=u_{h}}$, is the full
second order linearization and if it is inverted exactly then
Eq.~\ref{functional_error3} is identical to an adjoint method where the higher
order remaining error terms have been neglected and the adjoint
variables, $\lambda$, are given by:

\begin{equation}\label{approx_adjoint_var}
\lambda^{T} = \frac{\partial J_{h}}{\partial U}\bigg|_{U=u_{h}}
\left[ \frac{\partial L_{h}}{\partial U}\bigg|_{U=u_{h}} \right]^{-1}
\end{equation}

\noindent For functional error estimation, the Jacobian must
be inverted exactly in order to obtain an accurate functional error estimate.
In practice, this inversion is too expensive and Eq.~\ref{approx_adjoint_var}
is solved using a krylov subspace method such as GMRES~\cite{Saad1986}.
But, for functional-based adaptation, we only require an adaptation indicator
which flags areas of the domain for adaptation.
Therefore, since the adjoint variables only serve as an indicator, they
need not be as accurate as is required for functional error estimation.
To approximate the adjoint variables, rather than solve
Eq.~\ref{approx_adjoint_var} using an iterative method, the inverse of the
Jacobian can instead be replaced with an approximate inverse.
This not only saves computational resources, but also allows the inverse of the
Jacobian to be recycled for multiple functionals.
In this paper, we investigate using sparse approximate inverse preconditioning
techniques to generate an approximate inverse of the residual
Jacobian~\cite{Chow1998,Wang2009}.

\subsection{Grid Adaptation}

\begin{figure}[H]
  \centering
  \includegraphics[width=4.0in]{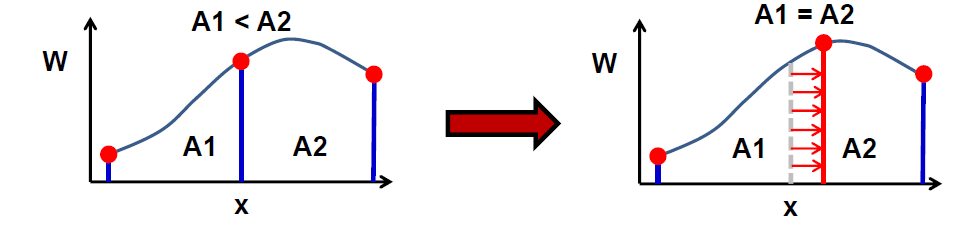}
  \caption{Weight Function Equidistribution~\cite{Jackson2015}}
  \label{equidistribution}
\end{figure}

Grid adaptation is performed using a structured adaptation module (SAM)
developed by Choudhary~\cite{Choudhary2013} which implements several 1D and 2D
r-adaptation schemes.
For this study, a center of mass adaptation procedure is used which
equidistributes a prescribed weight function across the computational domain.
During the equidistribution process, nodes of the mesh are shifted, as
illustrated in Fig.~\ref{equidistribution}, in order to
make the quantity cell size (i.e. $\Delta x$ in 1D) times weight function
equal-valued in all cells of the domain.

\noindent With this type of adaptation, the number of nodes in the mesh
remains constant throughout the adaptation process.
This can lead to significant computational savings over other forms of
adaptation, such as h-adaptation, where the number of nodes in the mesh can
increase.

Prior to conducting adaptation, care must be taken to formulate the weight
function in a manner which will achieve the desired adaptive behavior.
Ideally, it is preferred that resolution is increased in areas of the domain
where discretization error is large while decreasing resolution in areas where
it is small.
But, solely using discretization error to drive adaptation generally tends to
not achieve this goal as well as using truncation error to drive the adaptation
process~\cite{Roy2009,Zhang2000,Gu2001}.
While truncation error can be used to drive adaptation which targets the
overall discretization error in a solution, we seek a weight function which
adapts cells which contribute to the error in a functional.
Derlaga et al.~\cite{Derlaga2013,Derlaga2015a} and Derlaga~\cite{Derlaga2015b}
were able to accomplish this by weighting the local truncation error by the
adjoint sensitivities and, in some cases, were even able to achieve greater
reductions in functional error over other adjoint-based adaptation indicators.
In this study, the weight function proposed by
Derlaga et al.~\cite{Derlaga2013,Derlaga2015a} and Derlaga~\cite{Derlaga2015b}
is used to drive adaptation.
This weight function is formulated by normalizing the adjoint-weighted
truncation error in cell $i$ on a given mesh, $k$, by the infinity norm of the
adjoint-weighted truncation error on the initial mesh and averaging over
$N$ governing equations:

\begin{equation}\label{weight_function}
w_{i} = \frac{1}{N} \sum\limits_{j=1}^N
\frac{\bigl\lvert \lambda_{i,j} \tau_{h}(I_{h}^{q}u_{h})_{i,j}^{k} \bigr\rvert }{
\bigl\lvert\bigl\lvert \lambda_{j} \tau_{h}(I_{h}^{q}u_{h})_{j}^{k=1}
\bigr\rvert\bigr\rvert_{\infty} }
\end{equation}

\section{Test Case}

\subsection{Quasi-1D Nozzle}

The quasi-1D nozzle problem is used to investigate the effectiveness of the
proposed alternative to adjoint methods for functional-based adaptation.
The nozzle is characterized by a contraction which accelerates the flow to
Mach 1 at the throat followed by a diverging section which allows the flow to
expand subsonically or supersonically depending upon the pressure at the nozzle
exit.
Since the quasi-1D nozzle problem has an exact solution, it is the ideal test
bed for investigating these new methods for functional-based adaptation.

\subsubsection{Geometry \& Boundary Conditions}

\begin{figure}[H]
  \begin{minipage}[b]{0.45\linewidth}
    \centering
    \includegraphics[width=3.0in]{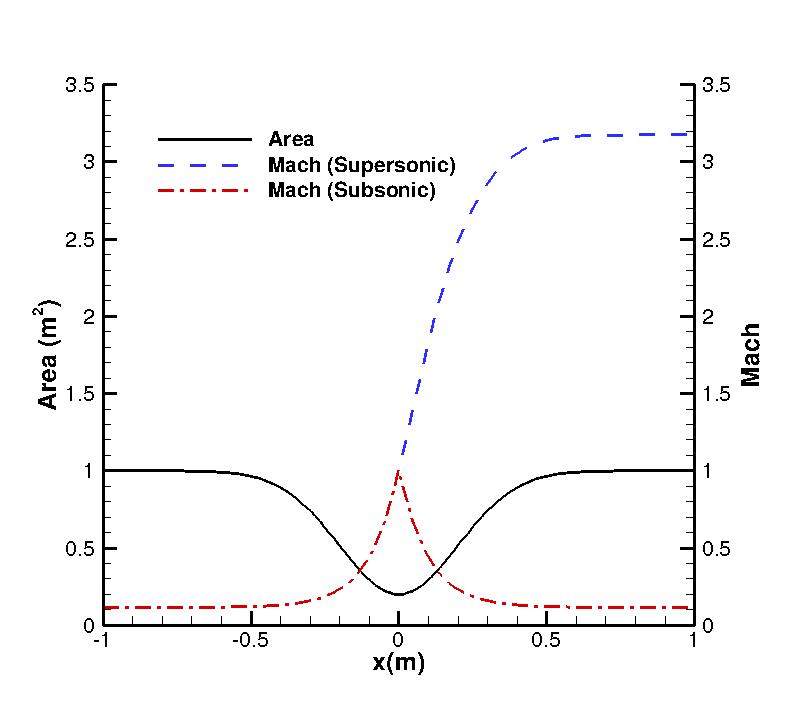}
    \caption{Gaussian Nozzle: Area/Mach Number Distribution}
    \label{gaussian_nozzle_area_mach}
  \end{minipage}
  \hspace{0.1\linewidth}
  \begin{minipage}[b]{0.45\linewidth}
    \centering
    \includegraphics[width=3.0in]{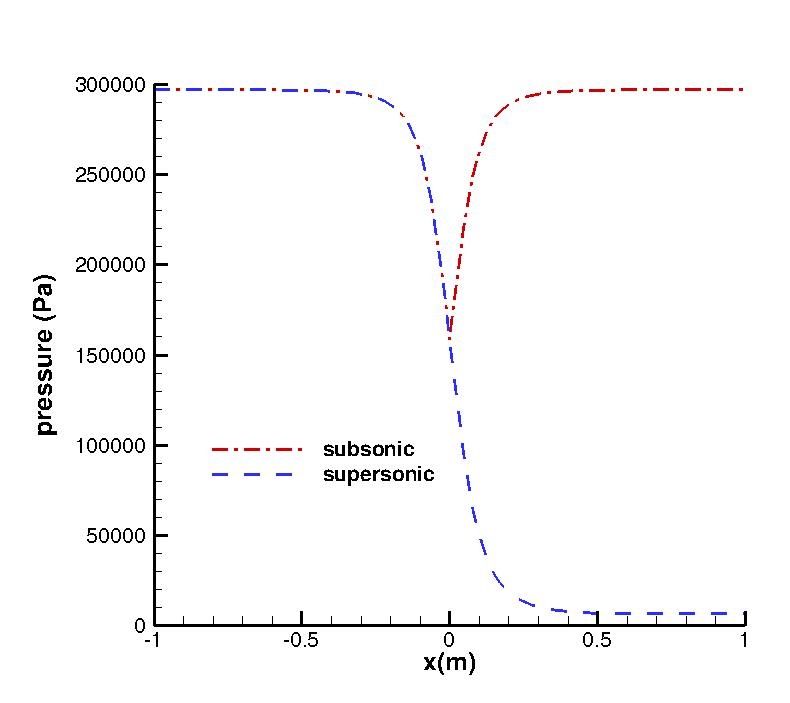}
    \caption{Gaussian Nozzle: Pressure Distribution}
    \label{gaussian_nozzle_pressure}
  \end{minipage}
\end{figure}

The geometry of the nozzle used in this study is taken from Jackson and
Roy~\cite{Jackson2015} and has a gaussian area distribution given by the
following:

\begin{equation}\label{nozzle_area}
A(x) = 1 - 0.8e^{\left( \frac{-x^{2}}{2\sigma^{2}} \right) }, \quad -1 \leq x \leq 1
\end{equation}

\noindent where $\sigma$ is chosen as 0.2.
This nozzle geometry was chosen for its lack of curvature discontinuities
which can cause spurious truncation error spikes~\cite{Jackson2015} and can make
truncation error estimation more diffucult.
For this study, only isentropic expansions are examined where the flow in the
diverging section is entirely subsonic or supersonic.
The area distribution and the Mach number distributions for both subsonic and
supersonic isentropic expansions can be seen in
Fig.~\ref{gaussian_nozzle_area_mach}.
The corresponding pressure distributions throughout the nozzle can be seen in
Fig.~\ref{gaussian_nozzle_pressure}.

At the inflow of the nozzle, the stagnation pressure and temperature are fixed
at 300 kPa and 600 K, respectively.
The Mach number is extrapolated from the interior to set the inflow state at
the face.
The outflow boundary conditions depend upon the expansion of the flow in the
diverging section of the nozzle.
For a supersonic isentropic expansion, all variables in the interior are
extrapolated to the face to set the outflow flux.
For a subsonic isentropic expansion, a back pressure is set and velocity and
density are extrapolated from the interior.
The back pressure is set to 297.158 kPa in this study.

\subsubsection{Numerical Methods: Finite Volume Discretization}

The flow in the nozzle is solved numerically using the quasi-1D form of the
Euler equations.
The Euler equations in weak, conservation form for a control volume fixed in
space take the following generic form:

\begin{equation}\label{euler_eq1}
\frac{\partial }{\partial t}\int_\Omega \vec{Q} \,d\Omega +
\oint_{\partial \Omega} \left( \vec{F}_{i} - \vec{F}_{v} \right) \,dS\ =
\int_\Omega \vec{S} \,d\Omega
\end{equation}

\noindent where $\vec{Q}$ is the vector of conserved variables, $\vec{F}_{i}$
is the vector of inviscid fluxes, $\vec{F}_{v}$ is the vector of viscous
fluxes, $\vec{S}$ is a generic source term, and $\Omega$ is the control
volume over which conservation of mass, momentum, and energy must be satisfied.
For the quasi-1D form of the Euler equations, the vector of conserved
variables, $\vec{Q}$, the fluxes, $\vec{F}_{i}$ and $\vec{F}_{v}$, and the
source term, $\vec{S}$, are given by:

\begin{equation}\label{euler_eq2}
\vec{Q} = \begin{bmatrix} \rho \\ \rho u \\ \rho e_{t}
\end{bmatrix}, \quad
\vec{F}_{i} = \begin{bmatrix} \rho u \\ \rho u^{2} + p \\ \rho u h_{t}
\end{bmatrix}, \quad
\vec{F}_{v} = \begin{bmatrix} 0 \\ 0 \\ 0 \end{bmatrix}, \quad
\vec{S} = \begin{bmatrix} 0 \\ p \frac{dA}{dx} \\ 0 \end{bmatrix}
\end{equation}

\noindent where $\frac{dA}{dx}$ describes how the cross-sectional area of the
nozzle changes down its length.
The system of equations is closed using the equation of state for a perfect gas
such that the total energy, $e_{t}$, and total enthalpy, $h_{t}$, are given by:

\begin{equation}\label{total_energy_enthalpy}
e_{t} = \frac{p}{\rho(\gamma-1)} + \frac{u^{2}}{2}, \quad
h_{t} = \frac{\gamma p}{\rho(\gamma-1)} + \frac{u^{2}}{2}
\end{equation}

\noindent where $\gamma$ is the ratio of specific heats for a perfect gas
which for air is $\gamma = 1.4$.

The quasi-1D form of the Euler equations, Eqs.~\ref{euler_eq1}
and~\ref{euler_eq2}, are discretized using a second order, cell-centered
finite-volume discretization.
The inviscid fluxes are computed using Roe's flux difference splitting
scheme~\cite{Roe1997} and Van Leer's flux vector splitting
scheme~\cite{VanLeer1982}.
MUSCL extrapolation~\cite{VanLeer1979} is used to reconstruct the fluxes to
the face in order to obtain second order spatial accuracy.
The boundary conditions are weakly enforced through the fluxes.
Numerical solutions are marched in time to a steady-state using an implicit
time stepping scheme~\cite{Beam1976} given by the following:

\begin{equation}\label{implicit_q1d_euler1}
\left[ \frac{\Omega}{\Delta t} I + \frac{\partial \vec{R}}{\partial \vec{Q}}
\right]^{n} \Delta \vec{Q}^{n} = - \vec{R}^{n}
\end{equation}

\noindent where $\Omega$ is the volume of a given cell, $\Delta t$ is the time
step, $I$ is the identity matrix, $\frac{\partial \vec{R}}{\partial \vec{Q}}$
is the Jacobian matrix, $\Delta \vec{Q}^{n}$ is a forward difference of the
conserved variable vector given by
$\Delta \vec{Q}^{n} = \vec{Q}^{n+1} - \vec{Q}^{n}$, and $\vec{R}^{n}$ is the
residual evaluated at time level $n$.
For the code used in this study, only the primitive variables are stored,
therefore, a conversion matrix is added to convert the conserved variable
vector, $\vec{Q}$, to the primitive variable vector, $\vec{q}$, such that:

\begin{equation}\label{implicit_q1d_euler2}
\left[ \frac{\Omega}{\Delta t} \frac{\partial \vec{Q}}{\partial \vec{q}} +
\frac{\partial \vec{R}}{\partial \vec{q}}
\right]^{n} \Delta \vec{q}^{n} = - \vec{R}^{n}
\end{equation}

\noindent where $\frac{\partial \vec{Q}}{\partial \vec{q}}$ is the conversion
matrix and $\vec{q}$ is the primitive variable vector given by
$\vec{q} = \left[ \rho, u, p \right]^{T}$.
The residual in a given cell, $\vec{R}_{i}$, for the quasi-1D Euler equations
is given by:

\begin{equation}\label{q1d_residual}
\vec{R}_{i}^{n} =
\vec{F}_{i+1/2}^{n} A_{i+1/2} - \vec{F}_{i-1/2}^{n} A_{i-1/2} -
\vec{S}_{i}^{n} \Delta x_{i}
\end{equation}

\noindent where $\vec{F}_{i+1/2}$ and $\vec{F}_{i-1/2}$ are the inviscid fluxes
at the faces on each side of cell $i$, $A_{i+1/2}$ and $A_{i-1/2}$ are the
corresponding face areas, and $\Delta x_{i}$ is the size of cell $i$.
For the primal solver, the Jacobian matrix,
$\frac{\partial \vec{R}}{\partial \vec{Q}}$, does not need to
be the full second order linearization since the left hand side of
Eq.~\ref{implicit_q1d_euler2} is just used to drive the residual, the right
hand side of Eq.~\ref{implicit_q1d_euler2}, to zero.
But, to maintain second order spatial accuracy, the residual must be the second
order discretization.

The discretization of the adjoint problem is performed using the discretization
outlined in \cite{Nielsen2004} and is given by:

\begin{equation}\label{implicit_adjoint}
\left[ \frac{\Omega}{\Delta t} \frac{\partial \vec{Q}}{\partial \vec{q}} +
\frac{\partial \vec{R}}{\partial \vec{q}} \right]^{T} \Delta \vec{\lambda}^{n} =
 - \left( \frac{\partial J_{h}}{\partial \vec{q}} +
\left[ \frac{\partial \vec{R}}{\partial \vec{q}} \right]^{T} \vec{\lambda}^{n}
\right)
\end{equation}

\noindent where $\Delta \vec{\lambda}^{n}$ is a forward difference of the
adjoint variables given by
$\Delta \vec{\lambda}^{n} = \vec{\lambda}^{n+1} - \vec{\lambda}^{n}$.
Similar to the primal solution, the Jacobian on the left hand side of
Eq.~\ref{implicit_adjoint} need not be the full second order linearization.
But, in order to obtain a usable adjoint solution, the Jacobian on the
right hand side of Eq.~\ref{implicit_adjoint} is required to be the full
second order linearization.

\subsubsection{Functionals of Interest}

For this study, two discrete solution functionals will be examined.
The first functional, the integrated value of pressure along the nozzle, is the
same functional used by Venditti and Darmofal~\cite{Venditti2000} for their
original work and is given by:

\begin{equation}\label{pressure_functional}
J_{h} = \int_{-1}^{1} p(x) dx
\end{equation}

\noindent The second functional under consideration is the integrated value of
entropy along the nozzle examined by Derlaga et al. ~\cite{Derlaga2013}:

\begin{equation}\label{entropy_functional}
J_{h} = \int_{-1}^{1} \frac{p}{\rho^{\gamma}} dx
\end{equation}

\section{Progress \& Future Work}

\begin{figure}[H]
  \begin{minipage}[b]{0.45\linewidth}
    \centering
    \includegraphics[width=3.0in]{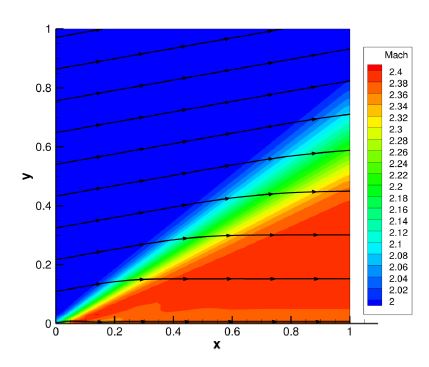}
    \caption{Expansion Fan: Mach Number Distribution~\cite{Derlaga2015b}}
    \label{expansion_fan_mach}
  \end{minipage}
  \hspace{0.1\linewidth}
  \begin{minipage}[b]{0.45\linewidth}
    \centering
    \includegraphics[width=3.0in]{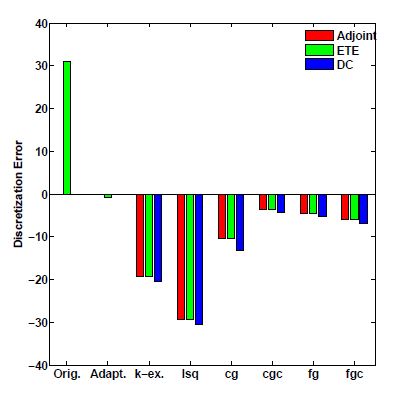}
    \caption{Force Functional Error Estimation~\cite{Derlaga2015b}}
    \label{expansion_fan_error_estimate}
  \end{minipage}
\end{figure}

The use of an adjoint method for functional-based error estimation and
adaptation can quickly become too costly when multiple functionals are of
interest.
Previous work from our research group has shown that local residual-based
error estimation can be used as a much cheaper alternative to adjoints when
multiple functional error estimates are required since local residual-based
error estimates can provide the same functional error estimate as an adjoint
method as long as the same linearization and truncation error estimate are
used~\cite{Derlaga2015b}.
An example of this from the work of Derlaga~\cite{Derlaga2015b} may be seen
in Fig.~\ref{expansion_fan_mach} and Fig.~\ref{expansion_fan_error_estimate}
for a 2D supersonic expansion fan where defect correction (DC), error transport
equations (ETE), and adjoints produced essentially the same error estimate for
the normal force on the wall.
For this paper, we now seek a more efficient method for functional-based
adaptation that will alleviate the need for an adjoint method when multiple
functionals are required.

Currently, both the primal solver and dual solver are implemented and running
properly for the quasi-1D nozzle problem due to the previous work of
Derlaga et al.~\cite{Derlaga2013} and Derlaga~\cite{Derlaga2015b}.
The structured adaptation module, SAM, is also implemented and interfaces well
with the quasi-1D nozzle code\cite{Derlaga2013,Jackson2015}.
As far as the solver, minimal changes will need to be made for the work
proposed in this paper.
Thanks to the hard work of previous developers, implementation of a sparse
approximate inverse for the Jacobian of the primal solution should be
relatively straightforward.
For the final draft of the paper, a sparse approximate inverse will be
implemented and weight functions for adaptation will be generated using the
methods outlined in this abstract.
These weight functions will be compared to those generated using a typical
adjoint method to examine the viability of the new approach for
functional-based adaptation.
The test results will include adaptation for both of the proposed functionals
for subsonic and supersonic cases of the quasi-1D nozzle.
Functional discretization error improvements will be examined for all test
cases using an adjoint method as well as the new, approximate adjoint approach
for functional-based adaptation.

\bibliography{Tyson_William_Aviation_2016}
\bibliographystyle{aiaa.bst}

\end{document}